\documentclass[twocolumn,showpacs,showkeys,preprintnumbers,superscriptaddress,amsmath,amssymb]{revtex4}
\usepackage[dvips]{graphicx}
\usepackage{latexsym,amssymb,amsmath,epsfig,bm,times,psfrag}
\usepackage{color}
\usepackage[bookmarksnumbered,bookmarksopen,colorlinks,citecolor=red,linkcolor=blue]{hyperref}
\usepackage{epstopdf}
\newcommand{\Fig}[1]{Fig.~\ref{#1}}
\newcommand{\eq}[1]{Eq.~(\ref{#1})}
\newcommand{\sect}[1]{Sec.~\ref{#1}}

\renewcommand{\part}{{\rm part}}

\newcommand{\be}{\begin{equation}}
\newcommand{\ee}{\end{equation}}
\newcommand{\bear}{\begin{eqnarray}}
\newcommand{\eear}{\end{eqnarray}}
\newcommand{\ba}{\begin{array}}
\newcommand{\ea}{\end{array}}

\begin{document}

\title{The role of spin polarization in event-by-event fluctuations}

\author{Xiao-Ying He}
\affiliation{School of Science, Guangxi University of Science and Technology, Liuzhou, 545006, China}
\author{Li-Juan Zhou}
\affiliation{Liuzhou Polytechnic University, Liuzhou, 545616, China}
\author{Li-Xin Yang}
\affiliation{School of Science, Guangxi University of Science and Technology, Liuzhou, 545006, China}
\author{Qian-Qian Du}
\affiliation{School of Physics and Technology, Guangxi Normal University, Guilin, 541004, China}
\author{De-Xian Wei}
\email{dexianwei@gxust.edu.cn}
\affiliation{School of Science, Guangxi University of Science and Technology, Liuzhou, 545006, China}
\date{\today}

\begin{abstract}
The event-by-event initial state fluctuations constitute a critical physical quantity for understanding the dynamics of high-energy heavy-ion collisions.
Due to the extremely limited available experimental data of such fluctuations, it is difficult to constrain or direct the characterization of the underlying initial state properties.
This work constrains the correlation function between the initial eccentricity and the asymmetry by employing a parton distribution function that incorporates spin polarization induced by thermal vorticity and thermal shear. Based on the kinetic theory calculations, we found that the asymmetry fluctuations considering spin polarization have a significant splitting effect compared to those without spin polarization.
These splitting exhibit pronounced system size and energy dependence.
Moreover, the ratios of correlation function display similar system size and energy dependence--particularly in the low-temperature regime ($T \lesssim$ 0.12 GeV).
These results suggest that spin polarization may serve as a useful constraint method for the event-by-event initial state fluctuations of heavy-ion collisions.

\end{abstract}



\maketitle

\section{Introduction}
\label{introduction}

High-energy heavy-ion collisions offer a unique laboratory for probing the properties of strongly interacting matter under extreme temperature and energy density.
Over the past few decades, experiments conducted at the Relativistic Heavy-Ion Collider (RHIC) and the Large Hadron Collider (LHC) have accumulated substantial empirical evidence supporting the formation of a deconfined quark-gluon plasma (QGP). 
Theoretical descriptions based on hydrodynamics~\cite{Heinz:2013th}
and microscopic transport approaches~\cite{Lin:2004en,Bass:1998ca} have demonstrated remarkable success in reproducing key observables from these collisions.
A critical advancement in recent years has been the recognition that event-by-event (EbE) fluctuations in the initial geometry must be incorporated to accurately describe two-particle correlation in the final state~\cite{Takahashi:2009na}, particularly odd-order anisotropic flow harmonics (e.g., triangular flow, $v_3$)~\cite{Alver:2010gr}.
Despite the profound theoretical importance of characterizing EbE initial state fluctuations in heavy-ion collisions, experimental constraints remain severely limited.
This is not surprising, as most measurements focus on the residual hadron matter at the final state of the collisions. Due to the fact that this system undergoes fluid dynamics evolution over a long period of time, the memory of its early stage almost completely disappears~\cite{Heinz:2013th,Luzum:2013yya}.
Consequently, initial states properties are typically inferred indirectly via two complementary strategies: (i) extending the set of hadronic observables simultaneously described by a given hydrodynamics with fixed initial conditions, and (ii) identifying observables that exhibit heightened sensitivity to initial state structure while maintaining robust insensitivity to later stage dynamical parameters, such as the specific shear viscosity ($\eta/s$) or the equation of state (EoS). 
Both approaches have yielded valuable insights into initial state granularity and spatial correlations.
Nevertheless, the first strategy faces inherent challenges: nontrivial final state correlations encoding initial state fluctuations necessitate phenomenological modeling that disentangles contributions from distinct microscopic origins, e.g., nucleonic position fluctuations, partonic degrees of freedom, and color charge density fluctuations. Quantifying the relative contributions of these mechanisms remains challenging.
In contrast, this work adopts the second strategy, focusing on observables whose sensitivity to initial state features is sensitivity and whose dependence on subsequent hydrodynamic evolution is insensitivity.

The observation of hyperon spin polarization (SP) in heavy-ion collisions has highlighted the importance of spin degrees of freedom in the dynamics of hot and dense QCD matter, opening a new direction for understanding collective spin phenomena in relativistic quantum fluids~\cite{STAR:2017ckg,STAR:2019erd,STAR:2021beb,ALICE:2021pzu}.
While current theoretical frameworks remain incomplete in describing local SP, most models successfully reproduce the global SP of $\Lambda$ hyperons--thereby offering valuable insights into the relevant properties of QCD matter~\cite{Liang:2004ph,Florkowski:2017ruc,Wei:2018zfb,Florkowski:2019qdp,Fukushima:2020ucl,Weickgenannt:2020aaf,Shi:2020htn,Florkowski:2021wvk,Speranza:2021bxf,Das:2022azr,Becattini:2024uha,Palermo:2024tza,Becattini:2025oyi}. Recently, the spin correlation function of $\Lambda$-$\Lambda$ pairs has been proposed as a novel probe of initial states spin density fluctuations~\cite{Giacalone:2025bgm}. 
At the experimental level, RHIC has reported the first high-precision measurement of spin correlations in hadron pairs~\cite{Vanek:2023oeo}, demonstrating the feasibility of such observables and placing quantitative constraints on spin transport mechanisms and nontrivial quantum correlations in QCD matter.

Based on the above studies, the present work investigates how SP can serve as a sensitive probe of EbE initial state fluctuations. The work is organized as follows: \sect{sec:method} outlines the kinetic theoretical framework incorporating SP dynamics and fluctuations of the asymmetric; \sect{sec:results} presents the quantitative analysis of how SP contributes to the correlation between the initial eccentricity and the momentum-tensor asymmetry; and \sect{sec:summary} provides a concise summary and discussion.

\section{Fluctuations in kinetic theory}
\label{sec:method}

\subsection{Participant eccentricity}
\label{subsec:eccentricity}

Let's start with the classic energy-momentum tensor,
\begin{eqnarray}\label{tens:101}
T^{\mu\nu}(\tau,r) &=& \int\frac{d^{3}p}{(2\pi)^{3}}\frac{p^{\mu}p^{\nu}}{p^{0}}f[r(\tau),p(\tau)]\,.
\end{eqnarray}
The factor $d^{3}p/[(2\pi)^{3} p^{0}]$ is the Lorentz invariant integration measure in momentum space, where $p^{0}=\sqrt{m^{2}+p^{2}}$ is the on-shell energy of the particle.

By evaluating the energy-momentum tensor, physical quantities can be extracted, such as participant eccentricity $\varepsilon_{p}$, which are of great interest in the early time of collisions.
The eccentricity for massive particles in kinetic theory are defined as
\begin{eqnarray}\label{eccen:101}
\varepsilon_{p} &=& \frac{\sqrt{([T^{xx}]+[T^{yy}])^{2}+4([T^{xy}])^{2}}}{[T^{xx}]+[T^{yy}]}\,.
\end{eqnarray}
where the results of $T^{\mu\nu}(\tau,r)$ are chosen proper time at $\tau=0.4$ fm.
Since the system had not reached the thermodynamic equilibrium state at an early time ($\tau=0.4$ fm), we assumed that the influence of SP on such stage could be neglected.

\subsection{Spin polarizations in kinetic theory}
\label{subsec:kinetic}

To highlight the contribution of spin polarization (SP), one employs kinetic theory to compute the energy-momentum tensor. Taking into account the spin measure, the SP-dependent energy-momentum tensor is defined as
\begin{eqnarray}\label{tens:102}
\tilde{T}^{\mu\nu}(\tau,r) &=& \frac{1}{2}\int\frac{d^{3}p}{(2\pi)^{3}}\frac{p^{\mu}p^{\nu}}{p^{0}}\int dS\tilde{f}[r(\tau),p(\tau),s(\tau)] \nonumber\\
&=& \int\frac{d^{3}p}{(2\pi)^{3}}\frac{p^{\mu}p^{\nu}}{p^{0}}f[r(\tau),p(\tau)]  \nonumber\\
&\times& \frac{1}{2}\sum_{m,n=1}^{2}\int \left[\delta_{mn}+\Pi[r(\tau),p(\tau)]\sigma_{mn}\right]dS   \nonumber\\
&=& \int\frac{d^{3}p}{(2\pi)^{3}}\frac{p^{\mu}p^{\nu}}{p^{0}}f[r(\tau),p(\tau)]  \nonumber\\
&\times& \frac{1}{2}\sum_{m=1}^{2}\int \left[\delta_{mm}+\Pi[r(\tau),p(\tau)]\sigma_{mm}\right]dS \,.
\end{eqnarray}
Here, the spin-dependent phase-space distribution function in near-equilibrium state, can be parametrized as~\cite{Weickgenannt:2020aaf,Becattini:2013fla,Florkowski:2019gio}
\begin{eqnarray}\label{disf:101}
& & \tilde{f}_{mn}[r(\tau),p(\tau),s(\tau)]   \nonumber\\
&=& f[r(\tau),p(\tau)] \times \left[\delta_{mn}+\Pi[r(\tau),p(\tau)]\sigma_{mn}\right]\,.
\end{eqnarray}
And, $m,n=1,2$ are spin indices, $f[r(\tau),p(\tau)]$ is the spinless phase-space distribution function, $s(\tau)$ denotes the spin degree of freedom in the phase-space distribution, $\sigma_{mn}$ is a three-vector composed of Pauli matrices, and $\Pi[r(\tau),p(\tau)]$ represents the spin-polarization three-vector.
The invariant spin measure is parametrized as~\citep{Florkowski:2018fap}
\begin{eqnarray}\label{tens:103}
& & \frac{1}{2}\sum_{m=1}^{2}\int \left[\delta_{mm}+\Pi[r(\tau),p(\tau)]\sigma_{mm}\right]dS   \nonumber\\
&\approx& 2-\frac{\, \sinh[\chi\,\Pi[r(\tau),p(\tau)]]}{\chi\,\Pi[r(\tau),p(\tau)]}
\end{eqnarray}
with
\begin{eqnarray}\label{tens:104}
& & \int \langle\cdot\cdot\cdot\rangle dS=\frac{M}{\pi\chi}\int \langle\cdot\cdot\cdot\rangle d^{4}s\delta(s\cdot s+\chi^{2})\delta(p\cdot s) \,,
\end{eqnarray}
where $\chi^{2}=s(s+1)= {3}/{4}$, $M$ is the particle's mass, the factor $1/2$ denotes the normalization of the sum of particles and antiparticles (This work does not distinguish between particles and antiparticles, as the chemical potential is not introduced in this framework.) When spin is treated as an independent variable, disregarding contributions from thermal vorticity and thermal shear, the result of integrating the spin measure is $\int dS= 2$~\cite{Florkowski:2018fap,Bhadury:2020cop,Drogosz:2024gzv}. The distribution function sums over spin-up and spin-down particles.

The most common approach in polarization studies assumes that the SP tensor is given by the total of the thermal vorticity $\varpi_{\rho\sigma}$ and thermal shear $\xi_{\rho\sigma}$~\cite{Becattini:2013fla,Arslan:2025tan,Florkowski:2021xvy,Banerjee:2024xnd}:
\begin{eqnarray}\label{ther:101}
\varpi_{\rho\sigma} &=& -\frac{1}{2}(\partial_{\rho}\beta_{\sigma}-\partial_{\sigma}\beta_{\rho}) \nonumber\\
&=& \frac{-1}{2T}(\partial_{\rho}u_{\sigma}-\partial_{\sigma}u_{\rho}) +\frac{1}{2T^{2}}(u_{\sigma}\partial_{\rho}T-u_{\rho}\partial_{\sigma}T)\,,
\end{eqnarray}
and
\begin{eqnarray}\label{ther:102}
\xi_{\rho\sigma} &=& \frac{1}{2}(\partial_{\rho}\beta_{\sigma}+\partial_{\sigma}\beta_{\rho}) \nonumber\\
&=& \frac{1}{2T}(\partial_{\rho}u_{\sigma}+\partial_{\sigma}u_{\rho}) -\frac{1}{2T^{2}}(u_{\sigma}\partial_{\rho}T+u_{\rho}\partial_{\sigma}T)\,,
\end{eqnarray}
where $\beta_{\rho}=u_{\rho}/T$ with $u$ denoting the fluid four-velocity and $T$ the temperature. The local temperature is defined as $T=(\pi^{2}\varepsilon/72)^{1/4}$~\cite{Zhang:2015tta}, and $\varepsilon={T}^{00}$ denotes the energy density.
Here, the thermal vorticity $\varpi_{\rho\sigma}$ and the thermal shear tensor $\xi_{\rho\sigma}$ correspond to the antisymmetric and symmetric parts of the fluid velocity gradient, respectively.
The combined contribution of thermal vorticity and thermal shearis expressed as~\cite{Becattini:2021suc}
\begin{eqnarray}\label{ther:103}
\Omega_{\rho\sigma}=\varpi_{\rho\sigma}+2\hat{\tau}\frac{p}{E}\xi_{\rho\sigma}\,,
\end{eqnarray}
where $p$ and $E$ denote the particle's momentum and energy, respectively, and $\hat{\tau}$ is the unit normal vector.
The second term's factor is derived by coupling the antisymmetric stress-energy tensor with $\xi$, as demonstrated in ref.~\cite{Becattini:2021suc}.

For the study of SP, the combined contribution vector is expressed as the sum of thermal vorticity and thermal shear tensors: $\Omega^{\mu}=-1/2\epsilon^{\mu\rho\sigma\tau}\Omega_{\rho\sigma}u_{\tau}$.
Under the assumption of small thermal vorticity or thermal shear within the local thermodynamic equilibrium, a linear relationship arises between the average spin polarization vector and the thermal vorticity plus thermal shear.
Specifically, for spin-$1/2$ particles, the polarization vector satisfies $\Pi^{\mu}\approx \Omega^{\mu}[2s(s+1)/3]$ (vorticity- and shear-induced polarization, VSIP).

\subsection{Fluctuations of asymmetry}
\label{subsec:fluctuation}

To simulate EbE fluctuations, we start with the characterization of the asymmetry, which is close to thermodynamic equilibrium.
For a comparison, the spinless and SP-dependent asymmetry for massive particles in kinetic theory are defined as
\begin{eqnarray}\label{asymm:101}
A &=& \frac{[T^{xx}]-[T^{yy}]}{[T^{xx}]+[T^{yy}]}  \nonumber\\
\tilde{A} &=& \frac{[\tilde{T}^{xx}]-[\tilde{T}^{yy}]}{[\tilde{T}^{xx}]+[\tilde{T}^{yy}]} \,.
\end{eqnarray}
where the spinless energy-momentum tensor $T^{\mu\nu}$ and SP-dependent energy-momentum tensor $\tilde{T}^{\mu\nu}$ are defined in \eq{tens:101} and \eq{tens:102}, respectively.
After obtaining the asymmetry, we can calculate the fluctuation signal of these asymmetry.
The spinless fluctuations can be quantitatively assessed through EbE variations in asymmetry, denoted as $\delta A$, where $\delta A=\langle A^{2}\rangle-\langle A\rangle^{2}$ represents the deviation from the event averaged quantity.
$\langle\ \cdot\rangle$ means that all particles are averaged in total events.
Similarly, the fluctuations of SP-dependent asymmetry is defined as $\delta \tilde{A}=\langle \tilde{A}^{2}\rangle-\langle \tilde{A}\rangle^{2}$.

\begin{figure*}[tp]
\begin{center}
\includegraphics[width=1.0\textwidth]{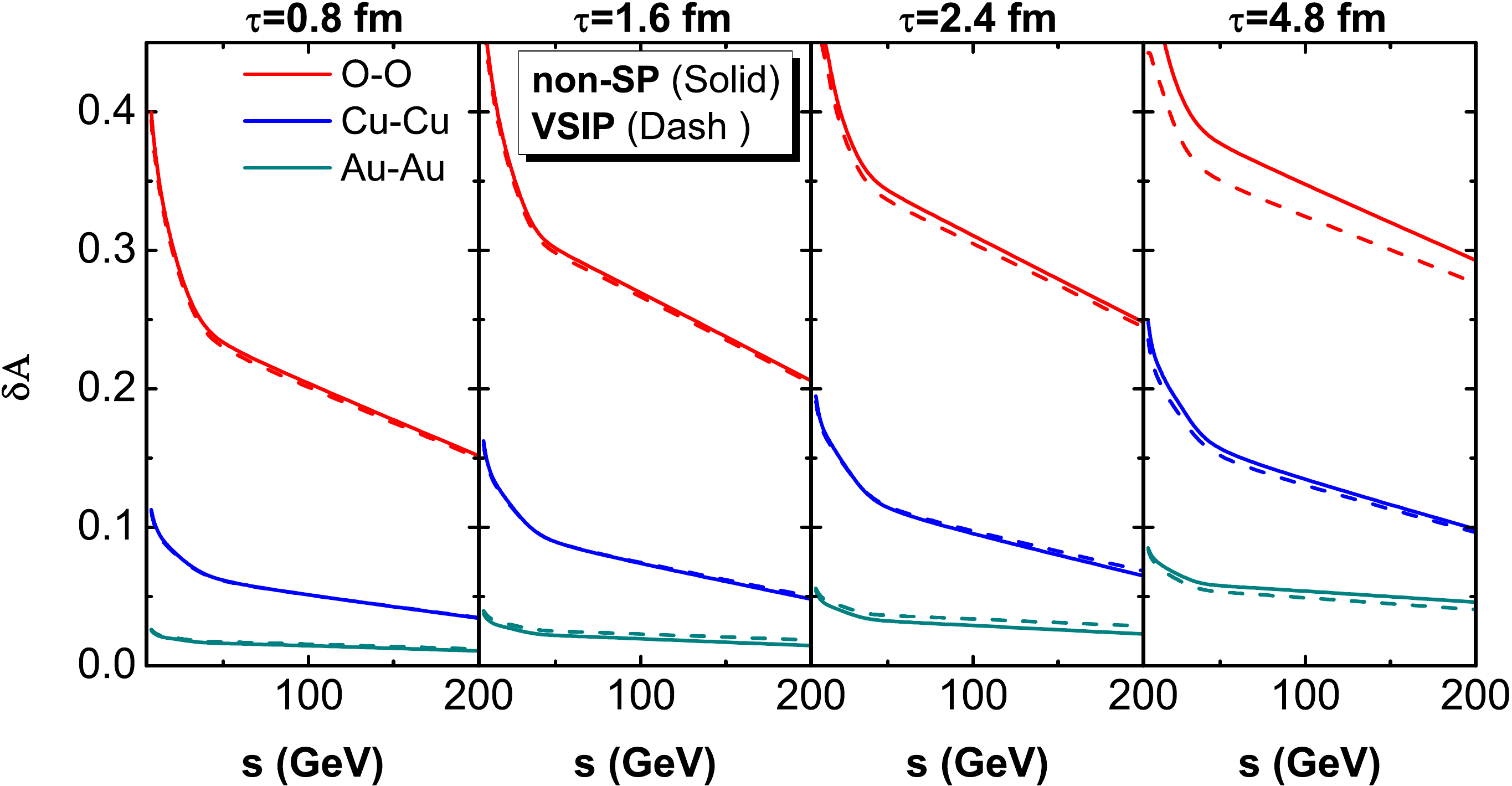}
\caption{(Color online)
Comparison of VSIP (dashed line) and non-SP (solid line) results: The fluctuations of asymmetry $\delta A$ as a function of energy for different systems.
All results are shown at proper time $\tau=0.8$, 1.6, 2.4, and 4.8 fm for each system.
}
\label{fig1}
\end{center}
\end{figure*}

In \Fig{fig1}, we show the results of comparison of $\delta A$ and $\delta \tilde{A}$ for different systems at different proper times.
The solid line represents the results of $\delta A$ (non-SP), while the dashed line represents the results of $\delta \tilde{A}$ (VSIP).
All results are shown at proper time $\tau=0.8$, 1.6, 2.4, and 4.8 fm for each system.
Both the $\delta \tilde{A}$ and the $\delta A$ are shown as functions of energy. From \Fig{fig1}, one can see that the splitting between $\delta \tilde{A}$ and $\delta A$ grows progressively with time.
Moreover, the splitting exhibits dependence on the system size, following the hierarchy: O $>$ Cu $>$ Au.
This is not surprising, where smaller systems contain fewer particles, leading to relatively larger fluctuations; conversely, larger systems feature higher particle densities and more frequent binary collisions, enabling faster attainment of chemical equilibrium. 
Interestingly, the contribution of SP to fluctuation magnitudes displays an inverted behavior: in O-O systems, SP exerts a suppressive effect on fluctuations, whereas in larger systems, its effect is non-monotonic--initially enhancing and subsequently suppressing fluctuations.
A primary origin of SP is the initial spin-orbital angular momentum coupling~\cite{Liang:2004ph,Becattini:2016gvu,Niida:2024ntm}.
This coupling retains information about the initial states.
A second major source is the time-dependent thermal shear.
During the early high-temperature phase, the dominant thermal vorticity suppresses both thermodynamic observables and fluctuation amplitudes in the QGP~\cite{Sahoo:2026cpn}.
While, beyond a characteristic evolutionary time, the thermal shear becomes dominant, exerting an effect opposite to that of the thermal vorticity~\cite{Florkowski:2021xvy,Becattini:2021iol}.
As illustrated in \Fig{fig1}, the non-monotonic temporal evolution of the ratio arises from this competition between the thermal vorticity and the thermal shear.

In this work, the EbE fluctuations are studied in O+O, Cu+Cu, and Au+Au collisions at 20-60 $\%$ centrality using a multi-phase transport (AMPT) model~\cite{Lin:2004en}. No multiplicity filter is applied; instead, the centrality is fixed, yielding a broad range of multiplicities from low to high. The default collision probability parameters in AMPT are used: $a=0.5$, $b=0.9$~GeV$^{-2}$, $\alpha_s = 0.33$, and $\mu = 3.2$~\citep{MacKay:2022uxo,Zhang:2023lzv}.

This work adopts natural units with $k_B=c=\hbar=1$.

\section{Numerical results and discussions}
\label{sec:results}

\begin{figure*}[tp]
\begin{center}
\includegraphics[width=0.96\textwidth]{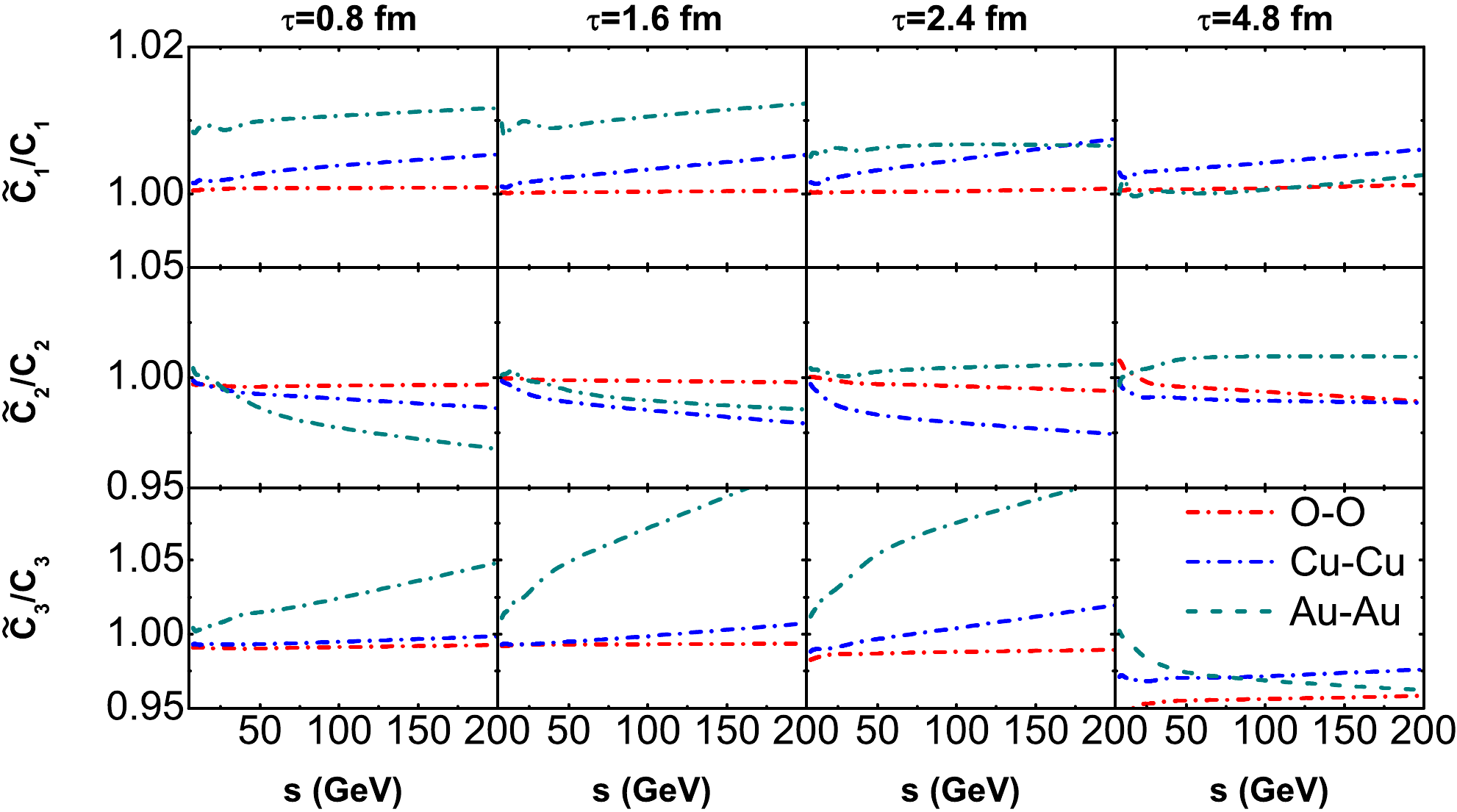}
\caption{(Color online)
Ratios of $\tilde{C}_{n}/C_{n} (n=1, 2, 3)$ as functions of energy for different systems.
All results are shown at proper time $\tau=0.8$, 1.6, 2.4, and 4.8 fm for each system.
}
\label{fig2}
\end{center}
\end{figure*}

\begin{figure*}[tp]
\begin{center}
\includegraphics[width=1.00\textwidth]{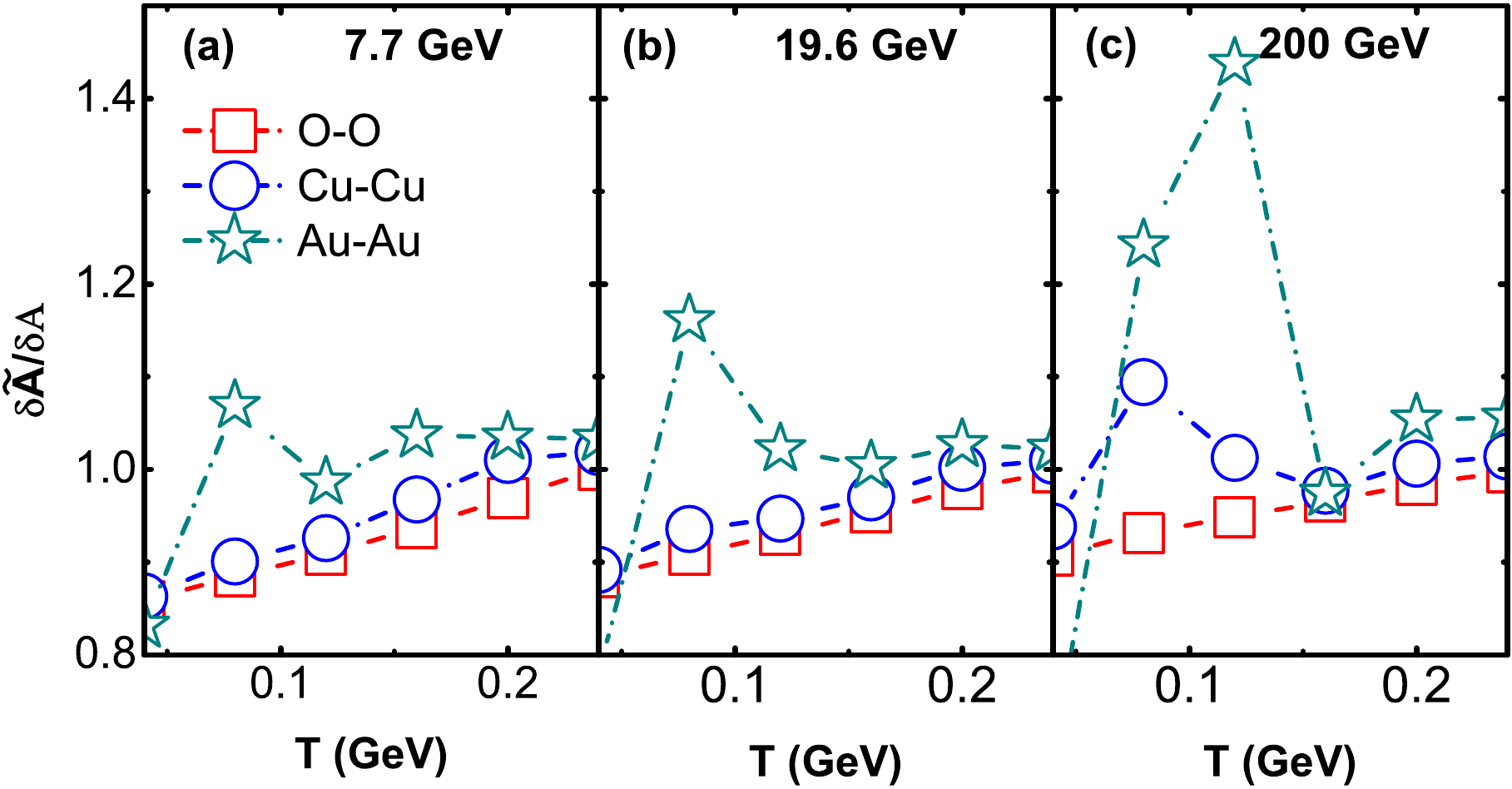}
\caption{(Color online)
Ratio of $\delta\tilde{A}/\delta A$ as functions of temperature for different systems on $\sqrt{s_{NN}}=7.7,~19.6$,~and 200 GeV.
All results are integrated over proper time from $\tau=0.8$ to 5 fm for each system.
}
\label{fig3}
\end{center}
\end{figure*}

\begin{figure*}[tp]
\begin{center}
\includegraphics[width=1.00\textwidth]{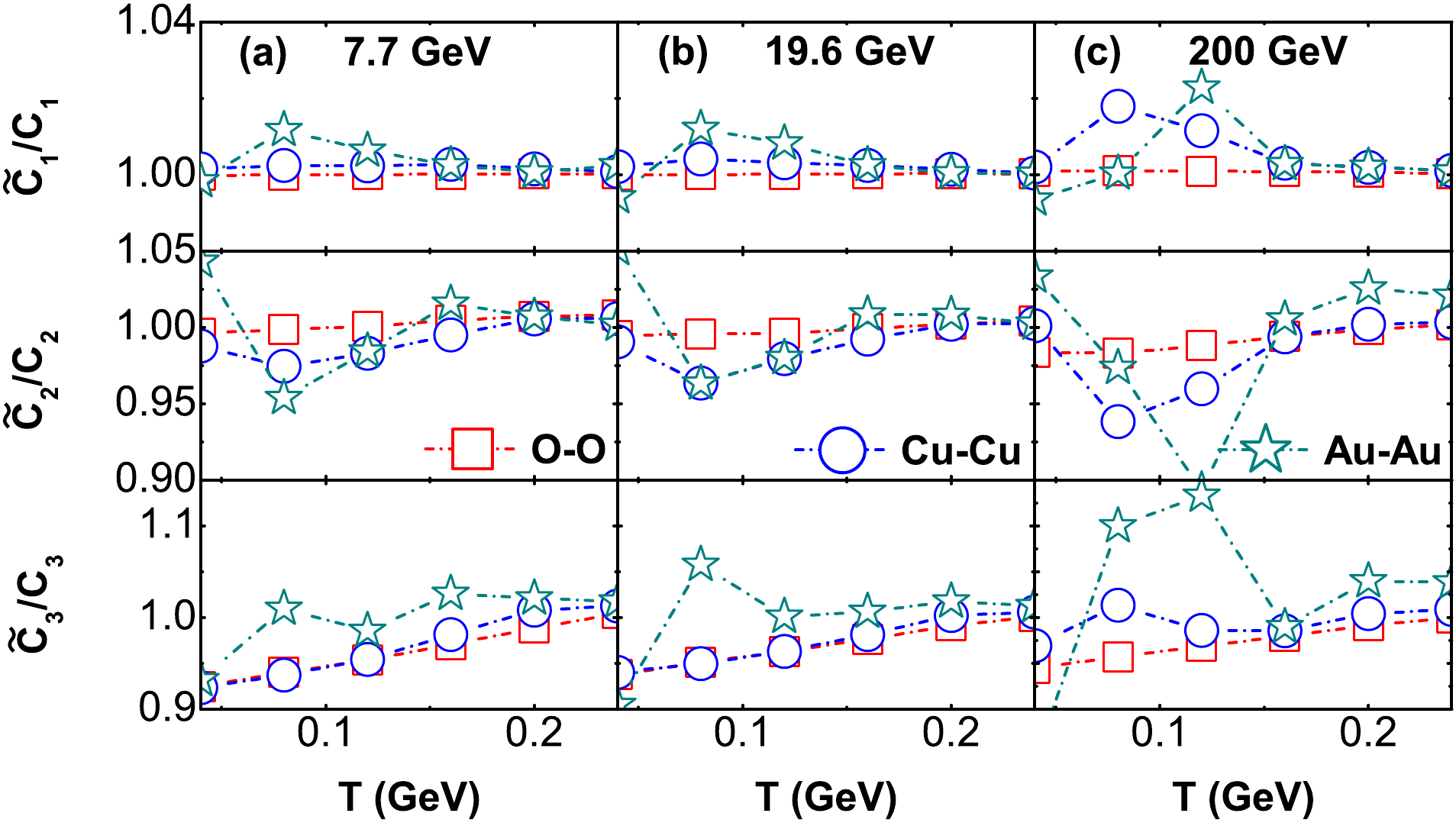}
\caption{(Color online)
Ratios of $\tilde{C}_{n}/C_{n} (n=1, 2, 3)$ as functions of temperature for different systems on $\sqrt{s_{NN}}=7.7,~19.6$,~and 200 GeV.
All results are integrated over proper time from $\tau=0.8$ to 5 fm for each system.
}
\label{fig4}
\end{center}
\end{figure*}

Having introduced the kinetic theory framework, one now presents numerical results for the correlations. A more detailed numerical analysis will be reported in a forthcoming study.

To characterize the significance of SP in the study of EbE initial state fluctuations, we extract the correlations between eccentricity and asymmetry. Three-type correlations in non-SP methods are introduced in this work
\begin{eqnarray}\label{corr:101}
C_{1} &=& \frac{\langle\varepsilon_{p}^{2}A^{2}\rangle-\langle\varepsilon_{p}^{2}\rangle\langle A^{2}\rangle}{\langle\varepsilon_{p}^{2}\rangle\langle A^{2}\rangle}  \nonumber\\
C_{2} &=& \frac{\langle\varepsilon_{p}^{2}\rangle\langle A^{2}\rangle}{\langle\varepsilon_{p}\rangle^{2}\langle A\rangle^{2}}  \nonumber\\
C_{3} &=& \frac{\langle\varepsilon_{p}\rangle\langle A\rangle}{\sqrt{\langle\varepsilon_{p}^{2}\rangle\langle A^{2}\rangle}} \,.
\end{eqnarray}

Similar correlations in VSIP methods are also introduced as
\begin{eqnarray}\label{corr:102}
\tilde{C}_{1} &=& \frac{\langle\varepsilon_{p}^{2}\tilde{A}^{2}\rangle-\langle\varepsilon_{p}^{2}\rangle\langle \tilde{A}^{2}\rangle}{\langle\varepsilon_{p}^{2}\rangle\langle \tilde{A}^{2}\rangle}  \nonumber\\
\tilde{C}_{2} &=& \frac{\langle\varepsilon_{p}^{2}\rangle\langle \tilde{A}^{2}\rangle}{\langle\varepsilon_{p}\rangle^{2}\langle \tilde{A}\rangle^{2}}  \nonumber\\
\tilde{C}_{3} &=& \frac{\langle\varepsilon_{p}\rangle\langle \tilde{A}\rangle}{\sqrt{\langle\varepsilon_{p}^{2}\rangle\langle \tilde{A}^{2}\rangle}} \,.
\end{eqnarray}
where, $C_{1}$ denotes the Pearson-type correlation function, while $C_{2}$ and $C_{3}$ represent the correlation functions of special ranks.

To examine the role of SP in phase-space distributions, we show the energy dependence of $\tilde{C}_{n}/C_{n} (n=1, 2, 3)$ for various systems at $\sqrt{s_{NN}} = 7$-200~GeV at different proper times in \Fig{fig2}.
As shown in Fig.~\ref{fig2}, the ratios of $\tilde{C}_{n}/C_{n}$ exhibit linear energy dependence.
In AMPT, hadronization proceeds continuously throughout the entire dynamic evolution. Consequently, as the system evolves, the thermal vorticity and SP gradually decrease with the system evolution~\cite{Jiang:2016woz,Karpenko:2016jyx,Wei:2018zfb}.
In contrast, the ratio of correlation function shows a significant dependence on system size: at early times, it scales linearly with system size; at later times, a distinct inflection point emerges, signaling a transition to non-monotonic scaling behavior. This behavior aligns consistently with predictions from asymmetry fluctuations, as illustrated in Fig.~\ref{fig1}.
From Fig.~\ref{fig2}, the thermal vorticity extends the lifetime of the QGP medium during hydrodynamic expansion~\cite{Sahoo:2023xnu,Wei:2025zsp,Wei:2025nvj}, implying that the weaker vorticity Au-Au systems undergo chemical freeze-out sooner than smaller systems~\cite{Alzhrani:2022dpi}.

For the relativistic nuclear collisions at very high energy, the chemical potentials are negligible, and temperature is the only effective intensive variable.  To characterize the temperature-dependent separation between VISP and non-SP, we show the temperature-dependent ratios in \Fig{fig3} and \Fig{fig4}.

In \Fig{fig3}, the ratio of $\delta \tilde{A}/\delta A$ as a function of temperature for different systems.
These ratios monotonically in the high-temperature regime ($T >$ 0.12 GeV), while show inconsistent results in the low-temperature regime: monotonic temperature dependence is observed in lighter O-O collisions or lower-energy Cu-Cu collisions, whereas non-monotonic dependence arises in heavier Au-Au or higher-energy Cu-Cu collisions.
In the early, high-temperature stage of evolution, SP acts to suppress EbE initial state fluctuations; stronger polarization yields greater suppression.
These results are consistent with the earlier research showing system size-dependent SP, namely $P(O) > P(Cu) > P(Au)$~\cite{Alzhrani:2022dpi}. In the late, low-temperature stage--dominated by contributions from the Cooper-Frye freeze-out hypersurface may modulate SP~\cite{Karpenko:2016jyx,Padhan:2026mwg}, potentially accounting for anomalous features in the results.
It is also worth noting that the temperature gradient constitutes a non-negligible contributor to SP~\cite{Florkowski:2021xvy,Arslan:2025tan}.
As evident from \eq{ther:101} and \eq{ther:102}, it can be seen that the temperature gradient components of the thermal vorticity and thermal shear play different roles, which leads to a competitive state when they are used to describe the time-dependent fluctuation behavior.

The ratio of $\tilde{C}_{n}/C_{n}$ as a function of temperature for different systems is shown in \Fig{fig4}.
These ratios exhibit only weak temperature dependence in the high-temperature regime ($T >$ 0.12 GeV), that is, the ratios tend to unity. In contrast, inconsistent results are observed in the low-temperature region. Specifically, for smaller O-O or lower-energy Cu-Cu collisions, the ratios remain weakly temperature dependent. While, for heavier Au-Au or higher-energy Cu-Cu collisions, the ratios display a non-monotonic dependence on temperature. These characteristics increase with the increase in collision energy.
These results are supported by \Fig{fig3}.

\section{Summary and discussions}
\label{sec:summary}

This work employs the AMPT model to simulate the space-time evolution of noncentral light- and heavy-ion collisions. Two sources of spin polarization (SP) are considered in the calculations: thermal vorticity and thermal shear.

In this work, we examine the role of spin polarization (SP) caused by thermal vorticity and thermal shear in describing the event-by-event initial state fluctuations of the relative relativistic heavy-ion collisions. Based on the kinetic theory calculations, we found that incorporating SP is essential for modeling fluctuations in asymmetry observables.

(I) System size effect: SP-corrected $\delta A$ and the correlation function exhibit pronounced dependence on system size. These results are due to the effects of thermal vorticity and thermal shear, which influence the lifetime of the medium~\cite{Sahoo:2023xnu,Wei:2025zsp,Wei:2025nvj}.
Moreover, the interplay between thermal shear and temperature gradient introduces additional complexity into the dynamical evolution of the medium.

(II) Collision energy effect: Both SP-corrected $\delta A$ and the correlation function show a weak dependence on collision energy.

(III) Temperature effect: Both SP-corrected $\delta A$ and the correlation function demonstrate strong overall temperature dependence. In the high temperature regime ($T >$ 0.12 GeV), their variation remains approximately unity.
While below this threshold temperature, there is a significant deviation.
Specifically, light O-O collisions and low-energy Cu-Cu collisions retain mild temperature dependence; while it displays non-monotonic temperature dependence in heavier Au-Au and higher-energy Cu-Cu collisions. The larger energy, the more significant the non-monotonic property becomes.
These behaviors are governed by competing physical mechanisms, including chemical freeze-out dynamics and the balance between thermal shear and temperature gradient effects~\cite{Fu:2021pok,Becattini:2021suc,Becattini:2021iol}.

Collectively, our research results suggest that SP as a sensitive and quantitative probe of event-by-event initial state fluctuations in heavy-ion collisions, offering a novel constraint for theoretical models of pre-equilibrium dynamics and early time symmetry breaking.
In the future, it will be essential to identify and extract observables that exhibit high sensitivity to SP while remaining robust against variations in other physical quantities. This is crucial for experimentally verifying the nature of the event-by-event initial state fluctuations.

\begin{acknowledgments}
This work is supported by Guangxi NSF (China) through Grant Nos.~2026GXNSFHA00640171, 2025GXNSFAA069552, and NSFC through Grant Nos.~12105057, 12505142.
\end{acknowledgments}

\bibliographystyle{apsrev4-1}
\bibliography{References}

\end{document}